\documentstyle[preprint,aps,floats,eqsecnum]{revtex}
\tighten

 \input amssym.def
 \input amssym.tex

\input BoxedEPS
\SetRokickiEPSFSpecial  
\HideDisplacementBoxes

\def\half{{\textstyle{1\over2}}}
\def\quar{{\textstyle{1\over4}}}
\def\fract#1#2{{\textstyle{\frac{#1}{#2}}}}
\def\beq{\begin{equation}}
\def\eeq{\end{equation}}
\def\bi{\begin{itemize}}
\def\ei{\end{itemize}}
\def\beqar{\begin{eqnarray}}
\def\eeqar{\end{eqnarray}}

\newcommand{\Ff}{{\cal F}}
\newcommand{\Gg}{{\cal G}}
\newcommand{\Hh}{{\cal H}}

\newcommand{\Pp}{{\cal P}}
\newcommand{\bcP}{\bbox{\cal P}}
\newcommand{\bB}{\bbox{\beta}}
\newcommand{\bW}{\bbox{\omega}}
\newcommand{\r}{{\bf r}}
\newcommand{\R}{{\bf R}}
\newcommand{\B}{{\bf B}}
\newcommand{\bL}{{\bf L}}
\newcommand{\f}{{\bf f}}

\newcommand{\n}{{\bf n}}
\newcommand{\bu}{{\bf u}}
\newcommand{\bv}{{\bf v}}

\newcommand{\G}{{\bf G}}
\newcommand{\q}{{\bf q}}
\newcommand{\bP}{{\bf P}}
\newcommand{\bp}{{\bf p}}
\newcommand{\bj}{{\bf j}}

\newcommand{\drh}{\dot{\rho}}

\newcommand{\rmd}{{\rm d\null}}
\newcommand{\rmdd}[1]{\rmd^d#1\,}
\def\boldnab{\bbox{\nabla}}

\def\boldphi{\bbox{\phi}}

\let\varkappa\kappa
\let\hat\widehat

\skip\footins = 14pt 
\begin{document}


\title{Fluid Dynamical Profiles and Constants of Motion \\
from d-Branes}

\author{R. Jackiw\footnotemark[1]}

\footnotetext[1] {\baselineskip=12pt This work is supported
in part by funds provided by  the U.S.~Department of Energy
(D.O.E.) under contract
\#DE-FC02-94ER40818.  \hfill MIT-CTP-2820 \hfill
hep-th/9902024 \hfill  February 1999}

\address{Center for Theoretical Physics\\ Massachusetts
Institute of Technology\\ Cambridge, MA ~02139--4307,
USA}

\author{A.P. Polychronakos}

\address{Theoretical Physics Department\\
Uppsala University, S-75108 Uppsala, Sweden\\ and\\
Physics Department,  University of Ioannina\\ 45110
Ioannina, Greece}

\maketitle
 \begin{abstract}%
Various fluid mechanical systems enjoy a hidden,
higher-dimensional dynamical Poincar\'e symmetry, which
arises owing to their descent from a Nambu-Goto action.
Also, for the same reason, there are equivalence
transformations between different models.  These
interconnections, summarized by the diagram below, are
discussed in our paper.

$${\BoxedEPSF{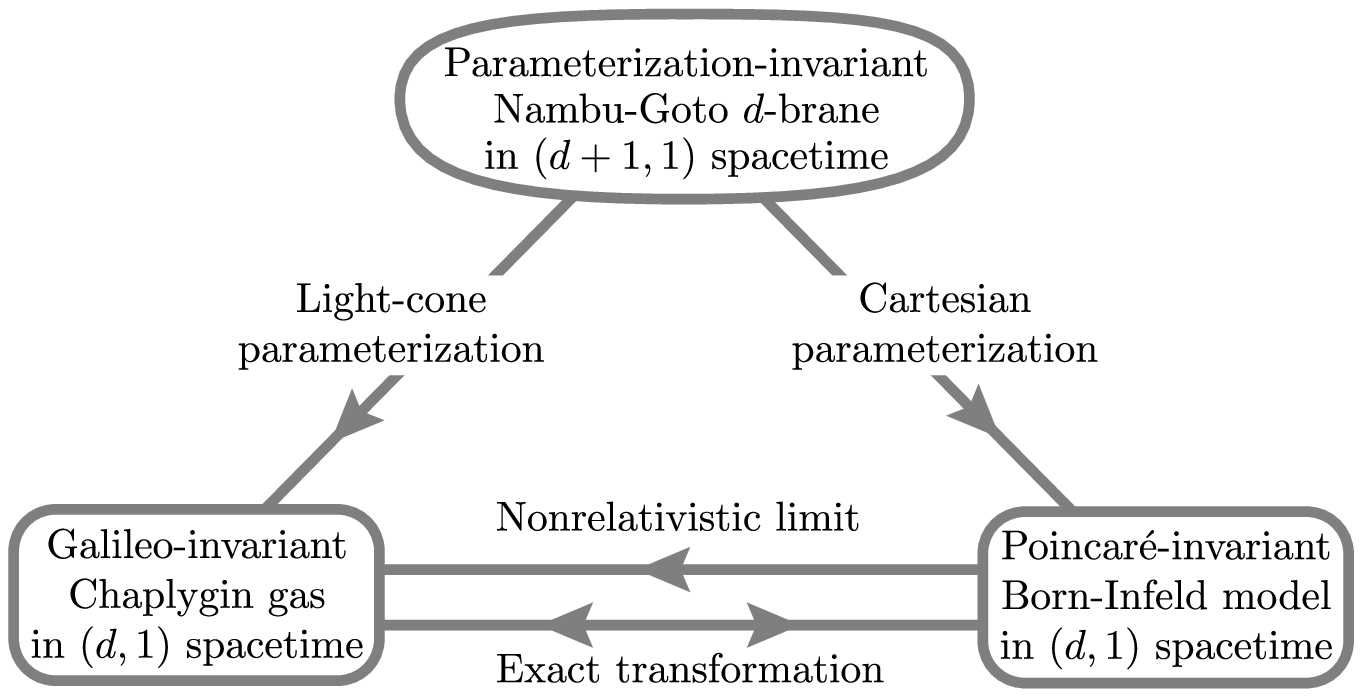 scaled 700}}$$
\end{abstract}
\newpage
\renewcommand{\baselinestretch}{1.2}

\section{Introduction}
\noindent 
In this paper we shall be concerned
with nonlinear dynamical systems that are described by a
density of ``matter'' $\rho$, flowing in time $\{ t\}$ with
local velocity
$\bv$ on a $d$-dimensional surface coordinated by $\{  \r
\}$.  The vectorial nature of $\bv$ is {\bf not\/}
unrestricted: $\bv$ is a function of $\boldnab
\theta$, where $\theta$ is a velocity ``potential'',  and we
shall examine several such functions. (When $\bv$ is linear
in $\boldnab \theta$, the flow is irrotational, $\boldnab
\times \bv=0$.)  The density is linked to the velocity by a
continuity equation involving the matter current $\bj = \bv
\rho$,
\beq
\dot{\rho} (t, \r) +\boldnab \cdot \Big(\bv  (t,\r)  \rho
(t,\r)\Big) = 0
\label{eq:2}
\eeq  while the velocity satisfies an ``Euler'' equation
\beq
\dot{\bv} (t, \r) +\bv (t,\r)
\cdot \boldnab  \bv  (t,\r) = \f (\rho, \bv)\ .
\label{eq:3}
\eeq  [The over-dot always indicates differentiation with
respect to the temporal argument, while the gradient
$\boldnab$ (unless further specified) differentiates the
spatial argument.]

We shall examine theories with various expressions for the
force (per unit mass)
$\f$, which lead to Galileo, Poincar\'e and additional
unexpected kinematical symmetries of the equations, and
which sometimes produce completely integrable systems,
with an infinite number of local conserved quantities.  The
existence of a velocity potential $\theta$ allows the above
equations to be formulated with an action principle, which is
usually unavailable for this purpose.  Consequently the
symmetries that we find are in fact Noether symmetries,
which leave the action invariant.

Additionally, we shall present limiting and  equivalence
transformations between different models, which allow
mapping solutions of one model onto solutions of another.

Various topics that we discuss have already appeared in the
literature.  A common feature  unites the diverse models
that we study: they have a common antecedent in that they
can be gotten from a parametrization-invariant Nambu-Goto
action for a d-brane on a
$d+1$-dimensional space, moving in $(d+1,1)$-dimensional
space-time.  [A ``d-brane'' is a $d$-dimensional extended
object: $d=1$ is a string, $d=2$ is a membrane, etc.  A
d-brane inhabiting $d+1$-dimensional space divides that
space in two.] When a light-cone parametrization  is selected
for the Nambu-Goto problem, one derives the Euler and
continuity equations for a
$d$-dimensional ``Chaplygin gas'' [in Eq.~(\ref{eq:3}), $\f
\propto \frac{1}{\rho^3}
\boldnab \rho$].  Alternatively, a Cartesian parametrization
produces the
$d$-dimensional ``Born-Infeld'' model (see below).

The relation between membranes and planar fluid
mechanics (the $d=2$ case) was known to Goldstone~\cite{ref:1},
and was developed by his student Hoppe (with 
collaboration)~\cite{ref:2}. Similar connections, yielding
equations in one spatial dimension, were discussed by Nutku (and
collaborators)~\cite{ref:3}.  Here we consider the general
$d$-dimensional case, and use the common ancestry of the
various fluid-mechanical models to posit unexpected
transformations between them, and to identify hidden,
dynamical symmetries in each model, which derive from the
high degree of symmetry of Nambu-Goto parent theory.

For $d=1$, the models that we study become especially
simple for two reasons.  First, their common antecedent is a
string (1-brane) moving on a plane (2-space), and for this
system the Nambu-Goto equations are integrable~\cite{ref:4}. 
Second, the requirement that velocity be expressible in terms of
a potential poses no restriction in one dimension, where any
function can be related to the derivative of another function. 
In this way one may understand that the
$d=1$ models are completely integrable, as has been noted
previously~\cite{ref:3,ref:4,ref:5}.

In Section II we consider noninteracting systems, with
Galileo- and Poincar\'e-invariant kinetic terms.  Specific
interactions that preserve Galileo and Poincar\'e symmetry,
as well as higher, dynamical symmetries are discussed in
Section III.  The relation of these to the Nambu-Goto theory
is explained in Section IV, where we also exhibit mappings
between the Galileo-invariant and the Poincar\'e-invariant
models.  The last Section V is devoted to models in one
spatial dimension.

\section{Force-Free Motion}
\noindent 
 The force-free problem, $\f=0$,
describes the free flow of dust.  Eqs.~(\ref{eq:2}) and
(\ref{eq:3}) are readily solved in terms of initial data,
specified (without loss of generality) at
$t=0$
\beqar
\rho (t, \r)\bigg|_{t=0} &=& \rho_0 (\r)
\label{eq:4} \\[1ex]
\bv (t, \r)\bigg|_{t=0} &=&\bv_0 (\r)\ .
\label{eq:5}
\eeqar Upon defining the retarded position $\q(t,\r)$ by the
equation
\beq
\q + t \bv_0 (\q) = \r
\label{eq:6}
\eeq  one verifies that (\ref{eq:2}) and (\ref{eq:3}), with
vanishing right side, are solved by
\beqar
\bv (t, \r)  &=&\bv_0 (\q )
\label{eq:7} \\
\rho (t, \r) &=&\rho _0 (\q ) \, \bigg| \det \frac{\partial
q^i}{\partial r^j}\bigg|\ .
\label{eq:8}
\eeqar

The free equations, with $\bv$ restricted to a function of
$\boldnab \theta$, possess a variational action formulation,
which was first given by Eckart for Galileo-invariant
nonrelativistic motion~\cite{ref:6}.  We reproduce and
generalize his argument.

The Lagrangian for $N$ point-particles of mass $m$ in free
nonrelativistic motion is the Galileo-invariant kinetic energy
$\half \sum^N_{n=1} m v_n^2 (t)$. In a continuum
description, the particle counting index $n$ becomes the
continuous variable $\r$, and the particles are distributed
with density $\rho$, so that
$\sum^N_{n=1}  v_n^2 (t)$ becomes $\int \rmdd{r} \rho (t,
\r) v^2 (t, \r)$.  But we also wish to link the density with the
current by the continuity equation (\ref{eq:2}), which can be
enforced with the help of a Lagrange multiplier $\theta$.
We thus arrive at the free continuum Lagrangian
\beq L^{\rm Galileo} =
 \int \rmdd{r} \Big[ \rho \half mv^2 + \theta
\Big(\dot{\rho} + \boldnab \cdot (\bv \rho)\Big)\Big]\ .
\label{eq:10}
\eeq  Since $L$ is first-order in time, and the canonical
1-form $
\int \rmdd{r}  \theta
\dot{\rho} \rmd t$ does not contain $\bv$, $\bv$ may be
varied, evaluated and eliminated~\cite{ref:7}.   We find
\beq
\rho  m \bv  - \rho \boldnab \theta  =0
\label{eq:11}
\eeq  showing that $\boldnab \theta$ is the local momentum
$\bp= m\bv$, and the velocity is irrotational:
$$
\bv =\frac{1}{m}  \boldnab \theta
$$
\beq
\boldnab \times \bv = 0\ .
\label{eq:12}
\eeq 
The Lagrangian (\ref{eq:10}) becomes
\beq L_0^{\rm Galileo} = \int \rmdd{r} \bigg[\theta
\dot{\rho} -
\rho \frac{(\boldnab
\theta)^2}{2m}\bigg]
\label{eq:13}
\eeq  where the subscript $0$ denotes absence of interaction.

Varying $\theta$ in  (\ref{eq:13}) regains the continuity
equation  (\ref{eq:2}), while varying $\rho$ produces the
free ``Bernoulli'' equation for the velocity potential
$\theta$ 
\beq
\dot{\theta} +\frac{1}{2m} (\boldnab \theta)^2 =0\ .
\label{eq:14}
\eeq  This is also recognized as the free Hamilton-Jacobi
equation. The  gradient of (\ref{eq:14}) gives rise, in view of
(\ref{eq:12}), to the free Euler equation (\ref{eq:3}) (with
$\f=0$).

Remarkably the same equations emerge for a kinetic energy
$T$ that is an arbitrary function of $\bv$.  If we generalize
(\ref{eq:10}) to
\beq L_0 = \int \rmdd{r} \Big[\rho T(\bv) +
\theta(\dot{\rho} + \boldnab \cdot (\bv \rho)
\Big]
\label{eq:15}
\eeq  we get, instead of (\ref{eq:11}),
\beq
\frac{\partial T(\bv)}{\partial \bv} \equiv \bp = \boldnab
\theta
\label{eq:16}
\eeq  and (\ref{eq:13}) becomes
\beq L_0 = \int \rmdd{r} \bigg[\theta \dot{\rho} -\rho
\bigg(\bv
\cdot \frac{\partial T (\bv)}{\partial \bv} - T(\bv)\bigg)
\bigg]
\label{eq:17}
\eeq  where it is understood that the Legrendre transform
of $T$ is expressed in terms of
$\boldnab \theta$ by inverting  (\ref{eq:16}).

Varying $\theta$ in (\ref{eq:17}) again gives the continuity
equation,
\beqar 0=\frac{\delta L_0}{\delta \theta} &=& \drh - \int
\rmdd{r}
\rho \bv \cdot
\frac{\delta}{\delta \theta} \bigg( \frac{\partial
T(\bv)}{\partial \bv}  \bigg)
\nonumber \\  &=& \drh - \int \rmdd{r} \rho \bv \cdot
\frac{\delta}{\delta \theta}
\boldnab
\theta\nonumber \\ &=& \drh + \boldnab \cdot (\bv \rho)
\label{eq:18}
\eeqar while varying $\rho$ leaves a generalization of the
free Bernoulli equation:
\beq 0=\frac{\delta L_0}{\delta \rho} =- \dot{\theta} - \bv
\cdot
\frac{\partial T}{\partial
\bv} + T(\bv)\ .
\label{eq:19}
\eeq  With the help of (\ref{eq:16}), this implies
\begin{mathletters}
\beqar
\frac{\partial}{\partial r^i}\dot{\theta} &=& -v^j
\frac{\partial^2 T}{\partial r^i
\partial v^j} = -v^j
\frac{\partial^2 \theta}{\partial r^i \partial r^j}  \nonumber
\\  &=& -v^j \frac{\partial^2 T}{\partial r^j \partial v^i} = -v^j
\frac{\partial v^k}{\partial r^j} \,
\frac{\partial^2 T}{\partial v^k \partial v^i}\ .
\label{eq:20a}
\eeqar 
On the other hand, from (\ref{eq:16}) it follows that
\beq
\frac{\partial}{\partial r^i}\dot{\theta} = \frac{\partial^2
T}{\partial v^i \partial v^k}
\dot{v}^k\ .
\label{eq:20b}
\eeq
\end{mathletters}%
Eqs.~(\ref{eq:20a}) and  (\ref{eq:20b})
are consistent only if  the free Euler equation (\ref{eq:3})
holds (provided the matrix $\frac{\partial^2 T}{\partial v^i
\partial v^j}$ has an inverse).

With a general form for $T(\bv)$, the local momentum
(\ref{eq:16}) $\bp \equiv
\frac{\partial T}{\partial {\bv}}$ remains
irrotational while the velocity, as determined by inverting
(\ref{eq:16}), becomes a nonlinear function of $\boldnab \theta$.

Evidently the solution  (\ref{eq:4})--(\ref{eq:8}) works with
arbitrary kinetic energy, whose specific form enters only in
the fixing the relation between $\bv$ and $\boldnab
\theta$.   However, the initial data for the velocity must be
consistent with the expression of the velocity in terms of
$\boldnab \theta$.

One may  present a family of constants of motion:
\beq C=\int \rmdd{r} \rho(t, \r) C\Bigl(\bv(t, \r), \r - t \bv
(t, \r)\Bigr)\ .
\label{eq:22}
\eeq  The time independence of $C$  is established either by
differentiating with respect to $t$ and using the free
equations of motion to prove that
$\frac{\rmd C}{\rmd t} =0$, or  more easily, by inserting the
solution  (\ref{eq:6})--(\ref{eq:8}) into  (\ref{eq:22}) and
changing integration variables from $\r$ to $\q$.  (Carrying
out these manipulations requires assuming that $\rho_0$
and $\bv_0$ obey appropriate regularity conditions and
drop off sufficiently at large distances.)

Various constants of motion arise from invariance against
time and space translation  (energy $E$ and total momentum
$\bP$, respectively) as well as space rotation (angular
momentum
$L_{ij}$), provided $T(\bv)$ carries no explicit time and
coordinate dependence, and does not depend on any external
vectors, i.e., $T(\bv) = T(v)$.  These constants are
\beqar E= H &=& \int \rmdd{r} {\Hh} \qquad {\Hh} = \rho
\Bigl(\bv \cdot \frac{\partial T}{\partial \bv} -
T(\bv)\Bigr)  \label{eq:24}
\\
\bP &=& \int \rmdd{r} {\bcP} \qquad {\bcP}= \rho \boldnab
\theta =  \rho \frac{\partial T (\bv)}{\partial \bv} = \rho
\bp
 \label{eq:25} \\  L_{ij}&=& \int \rmdd{r} (r^i {\Pp}^j - r^j
{\Pp}^i) \ .
\label{eq:26}
\eeqar 
Also shifting $\theta$ by a constant is a symmetry,
leading to conservation of
\beq N= \int \rmdd{r} \rho \ .
\label{eq:27}
\eeq  
To recognize that these constants of motion involve
particular forms for $C(\bv,
\r-t\bv )$ in (\ref{eq:22}), we recall that according to
(\ref{eq:16}) $\boldnab \theta$ is a function of $\bv$, and
the two are colinear when
$T(\bv) = T(v)$.

The densities $\Hh$ and ${\bcP}$ are components of an
energy-momentum tensor,
$T^{00}$ and $T^{0i}$ respectively, which satisfy continuity
equations with energy flux $T^{i0}$ and momentum flux
$T^{ij}$
\begin{mathletters}\label{eq:new27}
\begin{eqnarray} T^{00}&=&\Hh
\label{eq:new27a} \\ T^{i0}&=& v^i \Hh
\label{eq:new27b} \\ T^{0i}&=& \Pp^i
\label{eq:new27c} \\ T^{ij}&=&  v^i \Pp^j \ .
\label{eq:new27d}
\end{eqnarray}
\end{mathletters}
[$T^{ij}$ is symmetric when $T(\bv) =
T(v)$.] The continuity equations
\begin{mathletters}
\begin{eqnarray}
\dot{T}^{00} + \frac{\partial}{\partial r^i} T^{i0} &=&0
\label{eq:new28} \\
\dot{T}^{0j} + \frac{\partial}{\partial r^i} T^{ij} &=&0
\label{eq:new29}
\end{eqnarray}
\end{mathletters} are entirely equivalent to the free
dynamical equations (\ref{eq:2}) and  (\ref{eq:3}), with
vanishing force.

The symplectic structure, which is determined by the
canonical 1-form in (\ref{eq:17}), indicates that the only
nonvanishing bracket is \cite{ref:7}
\beqar
\{\theta (t,\r), \rho (t,\r') \} &=& \delta (\r - \r')
\label{eq:28}
\\
\noalign{\noindent  or equivalently}
\{\bp (t,\r), \rho (t,\r') \} &=& \boldnab \delta (\r - \r') \ .
\label{eq:29}
\eeqar  With these, one verifies that the constants of motion
(\ref{eq:24})--(\ref{eq:27}) generate the appropriate
infinitesimal transformation on the variables $\theta$ and
$\rho$.

Specific forms for $T(v)$ support additional, kinematical
symmetries and lead to further constants of motion.  In the
nonrelativistic case presented in
Eqs.~(\ref{eq:10})--(\ref{eq:14}), we have Galileo invariance
against boosts by velocity $\bu $.  The transformation law
for the fields reads
\beqar
\rho(t,\r) &\to& \rho_u (t, \r) = \rho (T, \R)
\nonumber\\[1ex]
\theta(t,\r) &\to& \theta_u (t, \r) = \theta (T, \R)+m(\bu
\cdot \r - u^2t/2)
\label{eq:36}
\eeqar 
where the transformed coordinates are boosted:
\beqar t \to T&=&t
\nonumber \\[1ex]
\r \to \R(t, \r)&=&\r - t \bu \ .
\label{eq:37}
\eeqar
 The inhomogenous terms in $\theta_u$ are
recognized as the well-known 1-cocycle of field theoretic
realizations of the Galileo group.  Also they ensure that the
transformation for the velocity $\bv = \boldnab \theta/m$
\beq
\bv(t,\r) \to \bv_u (t,\r) = \bv (t, \r- t\bu  ) + \bu
\label{eq:38}
\eeq is appropriate for the co-moving velocity.

The conserved quantity arising from the Galileo symmetry is
\beqar
\B&=& t\bP - m \int \rmdd{r} \r \rho
\nonumber \\ &=& -m \int \rmdd{r}  (\r - t \bv) \rho
\label{eq:39}
\eeqar where the last equality casts $\B$ in the form
(\ref{eq:22}).  With the help of the bracket (\ref{eq:28}),
$\B$ generates the infinitesimal transformation on the
fields, and its bracket with $\bP$ closes on $N$, thereby
exposing the familiar Galileo 2-cocycle, which provides an
extension of the algebra:
\beq
\{ B^i, P^j \} = \delta^{ij} m N \ .
\label{eq:40}
\eeq

The free Galileo-invariant theory possesses further
symmetries, which survive even in the presence of a
particular interaction.  Hence we postpone discussing them
until later, when interactions are included.

In the subsequent, in addition to the Galileo-invariant case,
we shall also be concerned with a relativistic,  Poincar\'e-invariant
model for which the point-particle kinetic energy is
$-mc^2
\sum^n_{n=1}
\sqrt{1-v^2_n (t)/c^2}$.  Upon passing to a continuum
description, as in the nonrelativistic case, we find
\beqar T(v) &=& -mc^2 \sqrt{1-v^2/c^2}
\label{eq:30} \\
\frac{\partial T(v)}{\partial \bv} &\equiv& \bp
=\frac{m\bv}{\sqrt{1-v^2/c^2}} =
\boldnab
\theta
\label{eq:31} \\
\bv &=& \frac{c \boldnab \theta}{\sqrt{m^2c^2 + (\boldnab
\theta)^2}}
\label{eq:32}
\eeqar leading to
\beq
\Hh = \rho c \sqrt{m^2c^2 + (\boldnab \theta)^2} =
\rho c \sqrt{m^2c^2 + p^2} = \rho
\frac{mc^2}{\sqrt{1-v^2/c^2}}
\label{eq:33}
\eeq  and Lagrangian
\beq L^{\rm Lorentz}_0 = \int \rmdd{r} \biggl[\theta \drh -
\rho c \sqrt{m^2c^2 + (\boldnab
\theta)^2} \, \biggr] \ .
\label{eq:34}
\eeq In the nonrelativistic limit this becomes
\beq L^{\rm Lorentz}_0 \to -mc^2N + L^{\rm Galileo}_0 \ .
\label{eq:35}
\eeq

Under Lorentz boosts, with velocity $\bu$, the fields
transform as
\beqar
\rho(t,\r)&\to& \rho_u (t,\r) = \rho(T, \R)
\frac{\dot{\theta} (T, \R) + c \sqrt{m^2c^2 + (\boldnab \theta
(T,\R))^2}} {\partial_t \theta (T, \R) +c \sqrt{m^2c^2 +
(\boldnab_{\r} \theta (T,\R))^2}}
\nonumber \\
\theta (t,\r)&\to&\theta_u (t,\r) = \theta (T, \R)
\label{eq:41}
\eeqar with Lorentz-boosted coordinates
\beqar t \to T(t, \r) &=& t \cosh \beta + \frac{1}{c}
\hat{\bB} \cdot
\r \sinh
\beta\nonumber
\\
\r \to \R(t, \r)&=& \r+\hat{\bB} \Big(ct \sinh  \beta +
\hat{\bB} \cdot \r (\cosh
\beta-1 )\Big)
 \label{eq:42}
\eeqar where $\bB=\bu/c$. Invariance is most easily
verified by writing the action corresponding to (\ref{eq:34})
as
\beq I^{\rm Lorentz}_0 =- \int \rmd t\, \rmdd{r} \rho \Bigl\{
\dot{\theta} + c\sqrt{m^2c^2 + (\boldnab \theta)^2}\,  \Bigr\} \ .
\label{eq:43}
\eeq The infinitesimal version of the field transformation
(\ref{eq:41})
\beqar
\delta \rho &=& \bB \cdot \Bigl( \frac{\r}{c}\,
\frac{\partial}{\partial t} + ct\boldnab \Bigr) \rho - \bB
\cdot
\frac{\bv}{c}\rho \nonumber \\
\delta \theta &=& \bB  \cdot \Bigl( \frac{\r}{c}\,
\frac{\partial}{\partial t} + ct\boldnab \Bigr)\theta
\label{eq:44}
\eeqar is generated by the Lorentz constant of motion
\beqar
\bL &=& t \bP - \int \rmdd{r} \r \Hh/c^2 \nonumber \\
&=& \int \rmdd{r}\Bigl(  t \rho\boldnab \theta -
\frac{\r}{c}\rho
\sqrt{m^2c^2 + (\boldnab \theta)^2}\,\Bigr)  \nonumber \\
&=& -m \int \rmdd{r}  (\r -  t\bv) \rho
\frac{1}{\sqrt{1-v^2/c^2}}
\label{eq:45}
\eeqar with the last equality exhibiting the form
(\ref{eq:22}). The transformation laws for $\rho$ and
$\theta$ ensure that
$j^\alpha =  (1,
\bv/c)\rho$ and $U^\alpha =
(1,\bv/c)\frac{1}{\sqrt{1-v^2/c^2}} $ transform as  Lorentz
vectors, so that
$\rho
\sqrt{1-v^2/c^2}$ and
$\theta$ are scalars~\cite{ref:8}.   The equation of motion
(\ref{eq:19}) for
$\dot{\theta}$ together with (\ref{eq:30})--(\ref{eq:32})
implies that the Lorentz vector
$\partial_\alpha \theta$ satisfies a Lorentz-invariant
equation
\beq (\partial_\alpha \theta)^2 = m^2c^2 \ .
\label{eq:45new}
\eeq

Note that the Lorentz transformation law (\ref{eq:41}) for
$\theta$ does not involve a 1-cocycle, which is a
nonrelativistic effect.  It is interesting to see how
(\ref{eq:36}) arises in the nonrelativistic limit.  By
comparing the relativistic action (\ref{eq:43}) to the
nonrelativistic one, we see that the relationship between
$\theta_{\rm R}$ and
$\theta_{\rm NR}$ --- the relativistic and nonrelativistic
variables --- is
\begin{equation}
\theta_{\rm R} = \theta_{\rm NR}-mc^2t \ .
\label{eq:46}
\end{equation} 
Applying the Lorentz transformation law
(\ref{eq:41}) to
$\theta_{\rm R}$ implies that
$\theta_{\rm NR} (t,\r) -mc^2t \to \theta_{\rm NR} (T,\R) -
mc^2T
$ or $\theta_{\rm NR}(t,\r)   \to \theta_{\rm NR} (T,\R) +
mc^2(t-T)$. We evaluate the nonrelativistic limit of the last
quantity from (\ref{eq:42}) and find
$mc^2(t-T) \to m(\bu  \cdot \r - u^2t/2) $, which matches
the 1-cocycle in (\ref{eq:36}).

Similar to the Galileo-invariant theory, this Poincar\'e-invariant  model
possesses further symmetries, which we shall discuss below, when we
include an interaction that preserves them.

\section{Motion in the Presence of Interactions}

\subsection{Nonrelativistic motion}
\noindent 
Interactions that
preserve the Galileo symmetry of the free, nonrelativistic
motion can be included by adding a
$\theta$-independent potential $V(\rho)$ to the Lagrangian
(\ref{eq:13}):
\beq L_{V}^{\rm Galileo} = \int \rmdd{r} \bigg[\theta \drh -
\rho \frac{(\boldnab
\theta)^2}{2m} - V (\rho)\bigg] \ .
\label{eq:48}
\eeq With nonvanishing $V$, (\ref{eq:6})--(\ref{eq:8}) are no
longer solutions, and the quantity (\ref{eq:22}) with
arbitrary  $C(\bv, \r - t\bv)$ is no longer constant.  Of
course the Galileo generators (\ref{eq:24})--(\ref{eq:27})
and (\ref{eq:39}), with
\beq
\Hh = \rho \frac{(\boldnab \theta)^2}{2m} + V (\rho) 
\label{eq:49}
\eeq remain time independent.  The energy-momentum
tensor retains the form (\ref{eq:new27}), with $\Hh$ given
by (\ref{eq:49}); in $T^{i0}$, $\Hh$ of (\ref{eq:49}) is
diminished by
$V-\rho \frac{\partial V}{\partial \rho}$; $T^{0i}$ is
unchanged and  $T^{ij}$ acquires the addition $\delta^{ij}
\Big(V-\rho \frac{\partial V}{\partial \rho}\Big)$.

The dynamics implied by  (\ref{eq:48}) arise in diverse
physical contexts.  The most directly physical application is
to isentropic, irrotational motion  in fluid mechanics with the
``force'' $f(\rho) = -\frac{\partial V(\rho)}{\partial \rho}$
corresponding to the enthalpy and $\Bigl( \rho
\frac{\partial^2 V(\rho)}{\partial \rho^2}\Bigr)^{1/2}$ is the
speed of sound~\cite{ref:9}.  Alternatively one finds
(\ref{eq:48}) (with $V$ depending also on $\boldnab \rho$)
in the hydrodynamical formulation of quantum mechanics,
which emerges when the wave function is presented
as~\cite{ref:10}
\beq
\psi = \rho^{1/2} e^{i \theta/\hbar} \ .
\label{eq:50}
\eeq 
[In this context, the inhomogenous Galileo
transformation of
$\theta$ (\ref{eq:36}) corresponds to the familiar change of
phase in a wave function under Galileo boosts, while shifting
$\theta$ by a constant is just the phase-invariance of
quantum mechanics, which leads to probability conservation,
i.e., constant $N$ in  (\ref{eq:27}).]

But we shall be especially concerned with the case
\beq V(\rho) = \lambda/\rho
\label{eq:51}
\eeq which arises in the study of ``d-branes'' ---
$d$-dimensional extended objects --- moving on a
$(d+1)$-dimensional space, in $(d+1,1)$-dimensional
space-time, and
 descending from a Nambu-Goto action (see
Section~IV)~\cite{ref:1,ref:2}.  Furthermore, (\ref{eq:51}) arises
in the nonrelativistic limit of a Poincar\'e-invariant model
with interactions, which we shall also describe below.

The equations of this theory, which follow from
\beq L_\lambda^{\rm Galileo} = \int \rmdd{r} \bigg\{
\theta\drh - \rho \frac{(\boldnab
\theta)^2}{2m} - \frac{\lambda}{\rho} \bigg\}
\label{eq:52}
\eeq read in their Bernoulli form
\beq
\drh + \boldnab \cdot \bigg(\frac{\boldnab \theta}{m} \rho
\bigg)=0
\label{eq:53}
\eeq
\beq
\dot{\theta} + \frac{(\boldnab \theta)^2}{2m}
=\frac{\lambda}{\rho^2}
\label{eq:54}
\eeq while their Euler form is gotten by recalling that $\bv =
\boldnab\theta/m$ 
\beq
\drh + \boldnab \cdot (\bv \rho)=0
\label{eq:55}
\eeq
\beq
\dot{\bv} + \bv \cdot \boldnab \bv
=\frac{-2\lambda}{m\rho^3} \boldnab \rho \ .
\label{eq:56}
\eeq These are the equations for a ``Chaplygin gas.''

For this model there exist  further symmetry
transformations~\cite{ref:11}.  The action
$I_\lambda^{\rm Galileo} =  \int \rmd t L_\lambda^{\rm
Galileo}$ is invariant against a rescaling of time, with
parameter $\omega$, under which the fields change
according to
\beqar
\rho(t,\r) &\to& \rho_{\omega} (t, \r) = e^{-\omega}\rho
(e^\omega t, \r)
\nonumber\\
\theta(t,\r) &\to& \theta_{\omega} (t, \r) = e^{\omega}
\theta (e^\omega t, \r)
\label{eq:57}
\eeqar and the time-independent generator of the
infinitesimal transformation is
\beq D=tH - \int \rmdd{r} \rho \theta \ .
\label{eq:59}
\eeq Furthermore, the action is also invariant against a
field-dependent diffeomorphism, implicitly defined by
\beqar t &\to& T(t,\r)= t + \bW \cdot \r + \half \omega^2
\theta (T,
\R)/m
\nonumber\\
\r &\to& \R(t, \r) = \r + \bW \theta (T, \R)/m
\label{eq:60}
\eeqar where the transformed fields are
\beqar
\rho(t,\r) &\to& \rho_\omega (t, \r) = \rho (T, \R)
\frac{1}{|J|}
\nonumber\\
\theta(t,\r) &\to& \theta_\omega (t, \r) =  \theta (T, \R)
\label{eq:61}
\eeqar and $|J|$ is the Jacobian of the transformation.
\begin{equation} J=\det
\left(
\begin{array}{cc}
\displaystyle \frac{\partial T}{\partial t} & \displaystyle
\frac{\partial T}{\partial r^j}  \\[2ex]
\displaystyle \frac{\partial R^i}{\partial t} & \displaystyle
\frac{\partial R^j}{\partial r^j}
\end{array}
\right) = \Big(1-\frac{\bW}{m} \cdot \boldnab \theta (T, \R)
- \frac{\omega^2}{2m}
\dot{\theta} (T,
\R)\Big)^{-1} \ .
\label{eq:62}
\end{equation} Here $\bW$ is a (vectorial) parameter of the
transformation, with dimension of (velocity)$^{-1}$, and the
time-independent quantity
\beq
\G = \int \rmdd{r} ( \r \Hh - \theta \bcP/m)
\label{eq:63}
\eeq generates the infinitesimal transformation.

Note that the generators $D$ and $\G$ remain
time-independent even in absence of the interaction
(\ref{eq:51}), hence these symmetries are also present for
the free theory.  The generators are {\it not\/} of the form
(\ref{eq:22}): they involve
$\theta$, and cannot be written in terms of $\bv = \boldnab
\theta/m$.  Finally we note that bracketing of the additional
generators with the Galileo generators of the nonrelativistic
theory on a $(d,1)$-dimensional space-time  produces an
algebra which is isomorphic to the Poincar\'e group in
$(d+1, 1)$ dimensions,~\cite{ref:2,ref:12} under which $(t,
\theta, \r)$ transforms as a
$(d+2)$-Lorentz vector $X^\mu$ in light-cone components
$(t=X^+,
\theta=X^-)$~\cite{ref:4}.

Using (\ref{eq:54}), we may eliminate $\rho$, and describe
the model solely in terms of $\theta$, whose dynamics is
governed by the Lagrangian
\beq L_\lambda = -2\sqrt{\lambda} \int \rmdd{r}
\sqrt{\dot{\theta} + \frac{(\boldnab
\theta)^2}{2m}} \ .
\label{eq:64}
\eeq It is seen that the ``interaction strength'' $\lambda$ in
fact disappears from the equations of motion for $\theta$;
$\lambda$ serves only to normalize the Lagrangian.  In the
free theory it is not possible to achieve this compact
formulation.  Furthermore, the dynamical equations can be
summarized by an equation for $\theta$, which follows from
(\ref{eq:53}), once (\ref{eq:54}) is used to eliminate $\rho$,
or alternatively the equation is derived from (\ref{eq:64}) 
\beq
\frac{\partial}{\partial t} \, \frac{1}{\sqrt{\dot{\theta} +
\frac{\displaystyle
 (\boldnab
\theta)^2}{\displaystyle  2m}}} + \boldnab \cdot \left( \frac{
\boldnab
\theta/m}{ \sqrt{\dot{\theta} +
\frac{\displaystyle  (\boldnab
\theta)^2}{\displaystyle  2m}}}\right)=0 \ .
\label{eq:64plus}
\eeq In spite of their awkward appearance, (\ref{eq:64})
and (\ref{eq:64plus}) are Galileo invariant in $(d,1)$
space-time, and possess a hidden, nonlinearly realized
Poincar\'e symmetry in $(d+1,1)$ space-time (which is a
descendant of the symmetries of the Nambu-Goto action; see
Section~IV).

Apart from the intrinsic interest in this nonlinear realization
of a kinematical/dynamical Poincar\'e symmetry, which  is
provided by (\ref{eq:57})--(\ref{eq:63}) supplementing the
linearly realized Galileo symmetry, the new symmetries
have the useful consequence of generating new solutions to
the equations of motion (\ref{eq:53})--(\ref{eq:56}) from old
ones.  For example, the time-rescaling invariant, particular
solution
\beqar
\theta (t, \r) &=& -\frac{mr^2}{2(d-1)t}  \nonumber \\[1ex]
\rho (t, \r) &=& \sqrt{\frac{2\lambda}{md}} \, (d-1)
\frac{|t|}{r}
\label{eq:65}
\eeqar can be transformed by (\ref{eq:60})--(\ref{eq:62})
into new solutions
$\theta_\omega$ and $\rho_\omega$, which are very
different in character from (\ref{eq:65})~\cite{ref:11}.  Note
that in (\ref{eq:65}) we must have $d>1$ and
$\lambda>0$.

At $d=1$, we can obtain general, time-rescaling invariant
solutions.  With the Ansatz $\theta \propto  {1}/{t}$,
(\ref{eq:64plus}) leads to a second order differential
equation for the $x$-dependence of $\theta$.  Therefore
solutions involve two arbitrary constants, one of which fixes
the origin of $x$, and can be ignored.  The other, which we
call $k$, appears in two distinct families of solutions (which
are related by an imaginary shift of $x$):
\begin{mathletters}
\begin{eqnarray}
\theta(t,x) &=& -\frac{m}{2k^2t} \cosh^2 kx
\label{eq:65plusa} \\[1ex]
\theta(t,x) &=& \frac{m}{2k^2t} \sinh^2 kx \ .
\label{eq:65plusb}
\end{eqnarray}%
\label{eq:65plus}%
\end{mathletters}
\noindent%
For real $\theta$, $k$ must be real or
imaginary.  When a real $\rho$ is computed from
(\ref{eq:54}), we find that $k$ must be real for $\lambda>0$,
imaginary for
$\lambda<0$, and a nonsingular density exists only with
(\ref{eq:65plusa}) for
$\lambda>0$ 
\beq
\rho(t,x) = \sqrt{\frac{2\lambda}{m}} \, \frac{k|t|}{\cosh^2
kx} \ .
\label{eq:65plus2}
\eeq The current $j = \frac{\theta'}{m}\rho$ exhibits a kink
profile (derivation with respect to the single spatial variable
is indicated by a dash)
\beq j(t,x) = \mp \sqrt{\frac{2\lambda}{m}} \, \tanh kx
\label{eq:65plus3}
\eeq where the sign is fixed by the sign of $t$.

In the last section we shall further review the $d=1$  case.

Another interesting solution, which is essentially
one-dimensional, even though it exists in arbitrary spatial
dimension, is given by
\beq
\theta(t, \r) = \Theta (\hat{\n} \cdot \r) + m \bu \cdot \r -
\half mt \Bigl(u^2 - (\hat{\n}
\cdot
\bu)^2\Bigr) \ .
\label{eq:65plus4}
\eeq 
Here $\hat\n$ is a spatial unit vector, and $\bu$ is an
arbitrary vector with dimension of velocity, while $\Theta$
is an arbitrary function with static argument, which can be
boosted by the Galileo transformation (\ref{eq:36}). The
corresponding charge density is time-independent:
\beq
\rho(t,\r) = \frac{\sqrt{2\lambda/m}}{\hat{\n} \cdot \bu +
\Theta' (\hat{\n} \cdot
\r)/m} 
\label{eq:65plus5}
\eeq and the static current becomes
\beq
\bj (t,\r)= \sqrt{\frac{2\lambda}{m}} \biggl( \hat{\n} +
\frac{\bu -
\hat{\n} (\hat{\n}
\cdot \bu)}{\hat{\n} \cdot \bu + \Theta' (\hat{\n} \cdot
\r)/m} \biggr) \ .
\label{eq:65plus6}
\eeq

\subsection{Relativistic motion}
\noindent
We seek an interacting generalization of $L_0^{\rm
Lorentz}$, which preserves Poincar\'e invariance.  To find
this, proceed as follows.  Let
\beq L^{\rm Lorentz} = \int \rmdd{r} \{ \theta \drh - \Hh
(\rho, \bp) \}
\label{eq:66}
\eeq with $\bp$ given by (\ref{eq:31}) and $\Hh$ to be
determined.  The symplectic structure is as in the free
theory, hence the Poisson brackets retain the form
(\ref{eq:28}), (\ref{eq:29}).  We calculate the Poisson
bracket between two Hamiltonian densities; in one the fields
are evaluated at $\r$, in the other at~$\r'$:
\begin{mathletters}
\beq
\{ \Hh (\r), \Hh(\r')\} = \int \rmd{\r}^{''} \rmd{\r}^{'''}
\biggl\{ \frac{\delta \Hh (\r)}{\delta \bp (\r^{''})} \cdot
\boldnab \delta (\r^{''}-\r^{'''})
\frac{\delta \Hh (\r')}{\delta \rho (\r^{'''})} - \r
\leftrightarrow \r' \biggr\} \ .
\label{eq:67a}
\eeq 
We assume that $\Hh$ is a local function of $\bp$ and
$\rho$, so that the functional derivatives lead to ordinary
derivatives, and (\ref{eq:67a}) becomes
\beq
\{ \Hh (\r), \Hh(\r')\} = \bigg( \frac{\partial  \Hh
(\r)}{\partial \bp}
\frac{\partial  \Hh (\r)}{\partial \rho} + \frac{\partial  \Hh
(\r')}{\partial \bp}
\frac{\partial  \Hh (\r')}{\partial \rho} \bigg)
\cdot
\boldnab \delta (\r - \r') \ .
\label{eq:67b}
\eeq%
\label{eq:67}%
\end{mathletters}%
On the other hand, the Dirac-Schwinger
condition for Lorentz invariance states that the bracket
(\ref{eq:67}) should give rise to $c^2$ times the momentum
density
$\bcP$, which in this problem  is given in  (\ref{eq:25}) as
\beq
\bcP = \rho \bp \ .
\label{eq:68}
\eeq 
Rotational invariance requires that the $\bp$
dependence of $\Hh$ is only on the magnitude $p$.  Thus
we conclude that
\beq 4 \frac{\partial}{\partial p^2} \Hh
\frac{\partial}{\partial \rho^2} \Hh = c^2 \ .
\label{eq:69}
\eeq

While many forms for $\Hh$ can solve (\ref{eq:69}), we
take a solution that is relevant to the present context, i.e., it
generalizes in a simple manner the free Hamiltonian density
(\ref{eq:33}), leads to a theory that descends from the
Nambu-Goto action, and coincides in the nonrelativistic limit
with the Galileo-invariant Chaplygin gas model:
\beq
\Hh = \sqrt{\rho^2c^2+a^2}\sqrt{m^2c^2 + (\boldnab
\theta)^2}  =  \sqrt{\rho^2+a^2/c^2}  \,
\frac{mc^2}{\sqrt{1-v^2/c^2}} \ .
\label{eq:70}
\eeq Evidently the parameter $a$ measures strength of
``interaction''.

Thus an interacting, Poincar\'e-invariant theory is described
by the Lagrangian
\beq L_a^{\rm Lorentz} = \int \rmdd{r} \Bigl[\theta \drh -
\sqrt{\rho^2c^2 +a^2} \, \sqrt{m^2c^2 + (\boldnab \theta)^2}
\,\, \Bigr] \ .
\label{eq:71}
\eeq The corresponding conserved Lorentz generator takes
the same form as in the first equality of  (\ref{eq:45}), with
$\Hh$ given by  (\ref{eq:70}), and it generates the
infinitesimal transformation
\beqar
\delta \rho &=& \bB \cdot \Bigl( \frac{\r}{c}
\frac{\partial}{\partial t}+ct\boldnab
\Bigr)
\rho -
\bB \cdot  \frac{\bv}{c} \sqrt{\rho^2 + a^2/c^2}  \nonumber
\\[1ex]
\delta \theta &=& \bB \cdot  \Bigl( \frac{\r}{c}
\frac{\partial}{\partial t} +ct\boldnab \Bigr)
\theta \ .
\label{eq:72}
\eeqar 
The finite transformation law is gotten by iterating
(\ref{eq:72}), and the generalization of (\ref{eq:41}) becomes
\beqar
\rho(t,\r) &\to& \rho_{u} (t, \r) = \rho (T, \R) \half
(\Omega_+ + \Omega_-) +
\sqrt{\rho^2(T, \R)  + a^2/c^2}   \half (\Omega_+ - \Omega_-)
\nonumber\\[1ex]
\theta(t,\r) &\to& \theta_{u} (t, \r) = \theta (T, \R)
\label{eq:73}
\eeqar where the Lorentz transformed coordinates $(T, \R)$
are given in (\ref{eq:42}) and
\beq
\Omega_{\pm} \equiv \frac{\dot{\theta} (T, \R) \pm
c\sqrt{m^2c^2 + (\boldnab
\theta(T, \R))^2}}{\partial_t \theta (T, \R) \pm c\sqrt{m^2c^2
+ (\boldnab_{\r}
\theta(T, \R))^2}} \ .
\label{eq:74}
\eeq
 It follows that $j^\alpha = \Big(\rho,
\frac{\bv}{c} \sqrt{\rho^2 + a^2/c^2}\,\Big)$ and
$U^\alpha =   \Big( \frac{\rho}{\sqrt{\rho^2 + a^2/c^2}},
\frac{\bv}{c} \Big) \frac{1}{\sqrt{1-v^2/c^2}}$
are Lorentz vectors, while $\theta$ is a scalar.

The equations of motion are
\begin{mathletters}\label{eq:76}
\beqar
\drh + \boldnab \cdot \Bigl( \frac{c\boldnab
\theta}{\sqrt{m^2c^2+(\boldnab\theta)^2}}
\sqrt{\rho^2 + a^2/c^2}  \Bigr) &=& \drh + \boldnab \cdot
(\bv \sqrt{\rho^2 + a^2/c^2} ) = 0 \label{eq:76a}\\
\dot{\theta} + \rho
c\frac{\sqrt{m^2c^2+(\boldnab\theta)^2}}{\sqrt{\rho^2  +
a^2/c^2} } &=& \dot{\theta} + \frac{\rho}{\sqrt{\rho^2  +
a^2/c^2}}
\frac{mc^2}{\sqrt{1-v^2/c^2}} = 0
\label{eq:76b} \ .
\eeqar
\end{mathletters} 
Using (\ref{eq:76b}) to express $\rho$
in terms of $\theta$,
\beq
\rho=-\frac{a}{c^2} \, \frac{\partial_0
\theta}{\sqrt{m^2c^2-(\partial_\mu \theta)^2}}
\label{eq:76c} 
\eeq 
and substituting this in (\ref{eq:76a}) yields a second
order, Lorentz covariant equation for $\theta$.  That
equation may also be gotten by eliminating $\rho$ from
(\ref{eq:71}) and deriving a Lagrangian for $\theta$.
\beq L_\theta = -a \int \rmdd{r}
\sqrt{m^2c^2-(\partial_\alpha \theta)^2}\ .
\label{eq:77}
\eeq This is  the ``Born-Infeld'' Lagrangian, leading to the
equation of motion
\beq
\partial^\alpha \bigg( \frac{1}{\sqrt{m^2c^2-(\partial_\mu
\theta)^2}}
\partial_\alpha
\theta \bigg) = 0\ .
\label{eq:78}
\eeq

As in the nonrelativistic theory [see (\ref{eq:64})] the
possibility of expressing
$\rho$ in terms of $\theta$ requires presence of the
``interaction,'' $a \ne 0$, whose nonvanishing strength
disappears from the nonlinear, interacting equations for
$\theta$.    The energy-momentum tensor for the theory
(\ref{eq:71}) is Lorentz covariant, of second rank, and
symmetric. After eliminating
$\rho$ with (\ref{eq:76c}), the resulting expression,
depending solely on $\theta$, bears the usual relation to its
Lagrangian (\ref{eq:77}).

Since the interacting Lorentz-invariant model is a
descendant of the Nambu-Goto Lagrangian (see Section~IV), it
comes as no surprise that it too possesses additional
kinematic symmetries, whose generators supplement the
generators of the linearly realized Poincar\'e group in
$(d,1)$ dimensions to give a nonlinear realization of
dynamical Poincar\'e algebra in $(d+1, 1)$
dimensions~\cite{ref:13}.

The additional symmetry transformations, which leave
(\ref{eq:71}) or (\ref{eq:77}) invariant, involve a
field-dependent reparametrization of time, defined
implicitly by
\beq t \to T(t, \r) = \frac{t}{\cosh mc^2 \omega} +
\frac{\theta(T, \r)}{mc^2} \tanh mc^2
\omega
\label{eq:83}
\eeq under which the field transforms according to
\beq
\theta(t, \r) \to \theta_\omega (t, \r)= \frac{\theta(T,
\r)}{\cosh mc^2\omega} - mc^2 t \tanh mc^2 \omega\ .
\label{eq:84}
\eeq [We record only the action of the transformations on
$\theta$; their effect on
$\rho$ can be read off from (\ref{eq:76c}).] The infinitesimal
generator, which is time independent by virtue of the
equation of motion (\ref{eq:78}), is
\beq D = \int \rmdd{r} \Big(m^2c^4 t \rho + \theta
\sqrt{\rho^2c^2+a^2} \sqrt{m^2c^2 + (\boldnab \theta)^2}
\Big) = \int \rmdd{r} (m^2c^4 t \rho + \theta \Hh)
\label{eq:85}\ .
\eeq

A second class of invariances involves a reparametrization
of the spatial variable, implicitly defined by
\beq
\r \to \R(t, \r) = \r - \hat{\bW} \theta (t, \R) \frac{\tan
mc\omega}{mc} +
\hat{\bW}
\hat{\bW} \cdot \r \bigg( \frac{1-\cos mc\omega}{\cos
mc\omega} \bigg)
\label{eq:86}
\eeq
\beq
\theta (t, \r)  \to \theta_{\bW} (t, \r) =
\frac{\theta(t, \R) - mc\hat{\bW}\cdot \r \sin
mc\omega}{\cos mc\omega}
\label{eq:87}
\eeq Here $\bW$ is a vectorial parameter, $\hat{\bW} =
\bW/\omega,  \omega =
\sqrt{\bW^2}$.    The time-independent generator of the
infinitesimal transformation reads
\beq
\G =\int \rmdd{r} (m^2c^2 \r \rho + \theta \rho \boldnab
\theta)=  \int \rmdd{r} (m^2c^2
\r\rho +\theta \bcP)\ .
\label{eq:88}
\eeq

With the addition of $D$ and $\G$ to the previous
generators, the Poincar\'e algebra in $(d+1,1)$ dimension is
reconstructed, and the transformation laws (\ref{eq:83}),
(\ref{eq:84}), (\ref{eq:86}), (\ref{eq:87}) ensure that $(t, \r,
\theta)$ transforms as a
$(d+2)$-dimensional Lorentz vector (in Cartesian
components)~\cite{ref:2}.  Note that this symmetry also
holds in the free, $a=0$, theory.

Because of the extended symmetry, one can generate new
solutions from old ones since both $\theta_\omega$ and
$\theta_{\bW}$ in (\ref{eq:84}) and (\ref{eq:87}) solve the
equation of motion if $\theta$ does.

A remarkable fact is that the nonrelativistic limit of the
above relativistic and interacting model precisely
corresponds to the nonrelativistic interacting model
discussed previously.  This is easily seen from (\ref{eq:71}),
which in the limit gives (\ref{eq:48}), with the help of
(\ref{eq:46}).
\beqar L_a^{\rm Lorentz}  &\to& -\frac{\rmd}{\rmd t} \int
\rmdd{r} mc^2 t\rho +
\int \rmdd{r}
\bigg[\theta_{NR} \drh - \rho \frac{(\boldnab
\theta_{NR})^2}{2m} - \frac{a^2}{2m\rho}
\bigg]
\nonumber \\
 &=&  -mc^2N + L_{a^2/2m}^{\rm Galileo}\ .
\label{eq:89}
\eeqar Equivalently, when $\rho$ is eliminated, we have
from (\ref{eq:77}) and (\ref{eq:46})
\beq L_\theta \to -a \int \rmdd{r} \sqrt{2m\dot{\theta}_{NR} -
(\boldnab \theta_{NR})^2} = L_{\lambda = a^2/2m}\ .
\label{eq:90}
\eeq Correspondingly, the equation of motion (\ref{eq:78})
goes over into (\ref{eq:64plus}).

It is easy to exhibit solutions of the relativistic theory, which
reduce to solutions of the nonrelativistic equations that were
given previously.  The following profiles solve (\ref{eq:78}).
\beq
\theta(t,\r) = -mc \sqrt{c^2t^2 + r^2/(d-1)}\ .
\label{eq:91}
\eeq With (\ref{eq:46}), this reduces to (\ref{eq:65}). In one
dimension we have
\beq
\theta(t,x) = -mc \sqrt{c^2t^2 + \cosh^2 kx/k^2}
\label{eq:92}
\eeq reducing to (\ref{eq:65plusa}).  The relativistic analog
of the lineal solution (\ref{eq:65plus4}) is
\beq
\theta(t,\r) = \Theta(\hat{\n} \cdot \r)+ m\bu \cdot \r-
mct  \sqrt{c^2+u^2  -(\hat{\n} \cdot \bu)^2}\ .
\label{eq:93}
\eeq Note that the above profiles continue to solve
(\ref{eq:78}), even when the sign of the square root is
reversed; but then they no longer possess a nonrelativistic
limit.

Additionally, there exists an essentially relativistic, chiral
solution describing massless propagation in one direction:
$\theta$ can satisfy the wave equation
\begin{mathletters}\label{eq:79}
\beqar
\Box \theta &=& 0
\label{eq:79a}\\
\noalign{\noindent when}
 (\partial_\mu \theta)^2 &=& {\rm constant} 
\label{eq:79b}
\eeqar
\end{mathletters}
as, for example, with plane waves
\beq
\theta (t, \r) = f(\hat{\n} \cdot \r \pm ct)
\label{eq:80}
\eeq
where $(\partial_\mu \theta)^2$ vanishes.  Then $\rho$ reads
from (\ref{eq:76c})
\beq
\rho = \mp \frac{a}{mc^2} f'\ .
\label{eq:81}
\eeq

\section{Relation to Nambu-Goto Action}
\noindent 
The Nambu-Goto action for a $d$-brane in $(d+1)$ spatial
dimensions,  moving in time on $(d+1,1)$ Minkowski space is
\beq I_{\rm N\hbox{-}G} = -\int \rmd \phi^0 \rmd\phi^1
\cdots \rmd \phi^d \sqrt{G}
\label{eq:4.1}
\eeq where $G$ is $(-1)^d$ times the determinant of the
induced metric
\beq G_{\alpha\beta} = \frac{\partial X^\mu}{\partial
\phi^\alpha} \,
\frac{\partial X_\mu}{\partial \phi^\beta}\ .
\label{eq:4.2}
\eeq Here Greek letters, from the beginning of the alphabet,
label the quantities
$\phi^\alpha=(\phi^0, \boldphi)$, with which the d-brane
coordinate $X^\mu$ is parametrized; $\phi^0$ is the
evolution parameter, and $\boldphi = \{ \phi^a, a=1,
\dots, d\}$ are the fixed-time, spatial parameters.  These
d-brane coordinates carry a Greek-letter index from the
middle of the alphabet, with value $0$ for the temporal
coordinate $X^0$ and $m$ for the $d$-brane's $d+1$ spatial
coordinates.
${\bf X} = \{ X^m, m=1, \dots, d, d+1 \}$.

The action is invariant against reparametrizations of the
$\phi^\alpha$, and we make the parametrization choice that
the $d$ coordinates $X^m, m=1, \dots, d$, are given by
$\phi^m$, which we rename $r^m$. For the remaining
parameters we use one of two options, ``light-cone'' and
``Cartesian''.

In the light-cone option, we define
\beq X^\pm = \frac{1}{\sqrt{2}} (X^0 \pm X^{d+1})
\label{eq:4.2plus}
\eeq and for $X^+$ choose the parametrization $X^+ =
\sqrt{2\lambda m} \, \phi^0$; also we rename $X^+$ as $t$.
The remaining coordinate $X^-$, a function of $\phi^0 =
t/\sqrt{2\lambda m}$ and $\boldphi = \r$, is renamed
$\theta(t, \r)/m$.  Upon evaluating the determinant $G$, we
see that the Nambu-Goto action coincides with the action for
(\ref{eq:64})~\cite{ref:14}. This identity also explains the
higher symmetry noted in Eqs.~(\ref{eq:57}) and
(\ref{eq:61}).  Our choice of parametrization does not
interfere with invariance against the $(d+1, 1)$ Poincar\'e
group, which acts on
$X^\mu$.  In the chosen parametrization, the Poincar\'e
transformation acts nonlinearly, mixing coordinates $(t,\r)$
with the field $\theta$.  (However, the higher symmetry is
also enjoyed by the noninteracting theory,
$\lambda=0$, which is {\bf not} equivalent to the
Nambu-Goto model.)

For the second, Cartesian option the chosen parametrization
permits writing $X^0$, which is renamed $ct$, as
$amc\phi^0$, while the last coordinate, $X^{d+1}$, which is a
function of $\phi^0=t/am$ and $\boldphi = \r$, is called
$\theta(t, \r)/mc$.  With these choices the Nambu-Goto
action coincides with that for the Born-Infeld Lagrangian,
 Eq.~(\ref{eq:77}).  Again the higher dynamical symmetry,
described by Eqs.~(\ref{eq:83})--(\ref{eq:88}), is now
understood as the covariance of the Nambu-Goto variables
$X^\mu$ against $(d+1,1)$-dimensional Poincar\'e
transformations.  (But once again, the similar invariance of
the free theory, $a=0$, {\bf cannot} be related to properties
of a Nambu-Goto action.)

Since both the $(d,1)$-dimensional Galileo-invariant
Chaplygin gas equations and the  $(d,1)$-dimensional
Poincar\'e-invariant Born-Infeld equations correspond to
different parametrizations of the $(d+1,1)$ Nambu-Goto
action, there must be a transformation --- recognized as a
reparametrization --- that takes solutions of one model into
the other.  This transformation may be formulated as follows.

Given a solution $\theta_{NR} (t, \r)$ to (\ref{eq:64plus}),
we solve for $T(t, \r)$ from the equation
\beq
\frac{1}{\sqrt{2}} \biggl( T(t, \r) + \frac{1}{mc^2} \theta_{NR}
\Big(T(t,\r),\r\Big)
\biggr) = t\ .
\label{eq:4.3}
\eeq Then a solution $\theta_{R} (t, \r)$ to (\ref{eq:78}), is
given by
\begin{mathletters}\label{eq:4.4}
\beq
\frac{1}{\sqrt{2}} \bigg( T(t, \r) - \frac{1}{mc^2} \theta_{NR}
\Big(T(t,\r),\r\Big)
\bigg) = \frac{1}{mc^2}\theta_{R} (t, \r)
\label{eq:4.4a}
\eeq or
\beq
\theta_{R} (t, \r) = mc^2 \Big( \sqrt{2} T(t, \r) -t \Big)\ .
\label{eq:4.4b}
\eeq%
\end{mathletters}%
Indeed this mapping produces solutions
(\ref{eq:91}), (\ref{eq:92}) and (\ref{eq:93}) from
(\ref{eq:65}), (\ref{eq:65plusa}) and (\ref{eq:65plus4})
respectively. [In fact both signs of the square root are
obtained; also (\ref{eq:93}) emerges from (\ref{eq:65plus4})
only after a redefinition of
$\Theta$ and
$\bu$.]

Oppositely, given as solution $\theta_R (t, \r)$ to the
relativistic Born-Infeld equation (\ref{eq:78}), we
reparametrize by solving for $T(t,\r)$ from
\beq
\frac{1}{\sqrt{2}} \bigg(T (t,\r) + \frac{1}{mc^2}\theta_{R}
\Big(T(t,\r),\r\Big) \bigg)= t
\label{eq:4.5}
\eeq and then find $\theta_{NR} (t, \r)$ from
\begin{mathletters}
\beq
\frac{1}{\sqrt{2}} \bigg(T (t,\r) - \frac{1}{mc^2}
\theta_{R}\Big(T(t,\r),\r\Big)\bigg) =
\frac{1}{mc^2} \theta_{NR} (t,\r)
\label{eq:4.6a}
\eeq or
\beq
\theta_{NR} (t,\r) = mc^2 \Big(\sqrt{2} T(t, \r) -t \Big)\ .
\label{eq:4.6b}
\eeq%
\label{eq:4.6}%
\end{mathletters}%

The two transformations are collected in the statement
\begin{eqnarray}
\theta_{NR} (t,\r) &=& mc^2 (\sqrt{2} T -t) \nonumber \\[1ex]
\theta_{R} (T,\r) &=& mc^2 (\sqrt{2} t -T)
\label{eq:4.7}
\end{eqnarray} with the instruction that  obtaining an
expression for $\theta_{NR}$ in terms of
$\theta_{R}$, or vice-versa, requires that one of $T$ or $t$
be eliminated in favor of the other.

The interrelationships may be summarized by the following
diagram:
$$
\centerline{\BoxedEPSF{jackiwfig.eps}}
$$

It is striking that there exists a two-fold relationship
between the Chaplygin gas and the Born-Infeld model.  First,
there is the exact mapping, given in (\ref{eq:4.7}), of one
into the other.  Second, the nonrelativistic limit
(\ref{eq:89}), (\ref{eq:90}) of the latter produces the former.

\section{One-Dimensional Motion}
\noindent 
In one spatial dimension, the motion of these systems
simplifies and  they become integrable. We shall give a
self-contained demonstration of integrability and derive the
integrals of motion in a compact form, stressing the
connection between the relativistic and nonrelativistic case.

In one dimension the requirement that the local momentum
is irrotational poses no restriction. We shall use as phase
space variables the local momentum $p(t,x)$ and the
particle density $\rho(t,x)$. The equal-time Poisson bracket
(\ref{eq:29}) becomes
\beq
\{p (x) , \rho(x') \} = \delta' (x-x')\ .
\label{eq:189}
\eeq Note that $p$ and $\rho$ in this relation are on an
equal footing and are governed by a nonlocal canonical
1-form $\half \int \rmd x\, \rmd y \dot{\rho} (x) \epsilon (x-y) p (y)
\rmd t$, where $\epsilon$ is the antisymmetric step function.
We shall consider local integrals, that is, quantities of the
form
\beq F = \int \rmd x \Ff ( p(x) , \rho(x) )
\label{eq:190}
\eeq with $\Ff$ a local function of $p$ and $\rho$. The
Poisson bracket of two such integrals $F$ and $G$ is
calculated through (\ref{eq:189}) as
\beq
\{ F , G \} = -\int \rmd x \Bigl[ (\Ff_{\rho \rho} \Gg_p +
\Ff_{\rho p} \Gg_\rho ) \partial_x \rho + (\Ff_{\rho p}
\Gg_p +
\Ff_{p p} \Gg_\rho ) \partial_x p \Bigr]
\label{eq:191}
\eeq where we suppressed the dependence on $x$ and
subscripts indicate partial derivative. If the above integrand
is a total $x$-derivative then (with appropriate boundary
conditions) the integral will vanish and
$F$ and $G$ will be in involution. For this we need the
curl-free condition
\beq (\Ff_{\rho \rho} \Gg_p + \Ff_{\rho p} \Gg_\rho )_p =
(\Ff_{\rho p} \Gg_p + \Ff_{p p} \Gg_\rho )_\rho
\label{eq:192}
\eeq 
or, finally
\beq
\frac{\Ff_{\rho \rho}}{\Ff_{p p}} = \frac{\Gg_{\rho
\rho}}{\Gg_{p p}}\ .
\label{eq:193}
\eeq
 Choosing one of the integrals to be the Hamiltonian $H
= \int \rmd x \Hh$, the well-known condition
\beq
\frac{\Ff_{\rho \rho}}{\Ff_{p p}} = \frac{\Hh_{\rho
\rho}}{\Hh_{p p}}
\label{eq:194}
\eeq guarantees that $F$ is a constant of the motion. If we
recover a set of such integrals satisfying (\ref{eq:194}) then
they will obviously  also satisfy (\ref{eq:193}) among
themselves. Therefore, constants of  motion will
automatically be in involution.

For the nonrelativistic case the Hamiltonian density
(\ref{eq:49}), (\ref{eq:51}) is
\beq
\Hh = \rho \frac{p^2}{2m} + \frac{\lambda}{\rho}
\label{eq:195}
\eeq and therefore the integrals of motion are generated by
functions that satisfy
\beq
\rho^4 \Ff_{\rho \rho} = 2\lambda m \Ff_{p p}\ .
\label{eq:196}
\eeq This can readily be solved by separation of variables.
We prefer, however, simply to give its general solution in
terms of two arbitrary functions
 $f$ and $g$ of one variable:
\beq
\Ff = \rho f \biggl(p+\frac{\sqrt {2\lambda m}}{\rho}\biggr) +
\rho g\biggl( p-\frac{ \sqrt {2 \lambda m}}{\rho}\biggr)\ .
\label{eq:197}
\eeq We essentially get two infinite towers of integrals.
Choosing, e.g.,
$f(z) = z^n$, $g=0$ or $g(z) = z^n$, $f=0$ we get the integrals
\beq I_n^\pm = \int \rmd x\, \rho \biggl(p \pm \frac{\sqrt
{2\lambda m}}{\rho}\biggr)^n\ .
\label{eq:198}
\eeq All the integrals presented in \cite{ref:5},\cite{ref:16}
can be identified as linear combinations of the $I_n^\pm$.
As stated, the $I_n^\pm$ are all in involution and
demonstrate the complete integrability of the system. The
Hamiltonian, in particular, is included as $4H = I_2^+ + I_2^-$
and the total momentum as $2P = I_1^+ + I_1^-$.

The quantities
\beq R_\pm = p \pm \frac{\sqrt {2\lambda m}}{\rho}
\label{eq:199}
\eeq appearing above are known as Riemann coordinates.
The equations of motion for this system (continuity and
Euler) are summarized in terms of $R_\pm$:
\beq {\dot R}_\pm = -\frac{1}{m} R_\mp \;  R'_\pm\ .
\label{eq:100}
\eeq Although this formulation for the equations is known,
the relation to the constants of motion does not seem to
appear in the literature.

Another fluid system for which the equations of motion,
expressed in terms of Riemann coordinates, take a form
similar to (\ref{eq:100}) possesses a potential cubic in
$\rho$:
$V(\rho) = \ell \rho^3/3$.  This also arises in a collective,
semiclassical description of nonrelativistic free fermions,
where the cubic potential reproduces fermion
repulsion~\cite{ref:17}.  In this case, the Riemann
corrdinates read
\beq R_{\pm} = p \pm \sqrt{2\ell m}\, \rho
\label{eq:101}
\eeq and, in contrast to (\ref{eq:100}), they decouple in the
the equations of motion:
\beq
\dot{R}_{\pm} = -\frac{1}{m} R_{\pm} R'_{\pm}\ .
\label{eq:102}
\eeq 
Indeed it is seen that $R_\pm$ satisfy essentially the
free Euler equation [compare (\ref{eq:3}) and identify
$R_\pm$ with $v$].  Consequently (\ref{eq:102}) is solved
by analogs of (\ref{eq:5})--(\ref{eq:7}).

Both these examples are special limiting cases of a more
general system, with potential
\beq V(\rho) = \frac{\lambda \bigl(\rho+\frac{2}{3}
a\bigr)}{(\rho+a)^2} + A \rho + B\ .
\label{eq:103}
\eeq 
(The terms $A,B$ correspond to a dynamically trivial
part that does not alter the equations of motion.)  The
Chaplygin gas corresponds to $a=A=B=0$, while the cubic
potential is recovered in the limit
\beq
\lambda=\ell a^4 \qquad A= \fract{1}{3} \ell a^2\qquad B=-\fract{2}{3}
\ell a^3, a \to \infty\ .
\label{eq:103plus1}
\eeq
 The Riemann coordinates are
\beq R_{\pm} = p \pm \frac{\sqrt{2\lambda m}}{\rho+a}
\label{eq:103plus2}
\eeq and the conserved integrals of this system are given by
functions of $R_{\pm}$:
\beq {\cal F} = (\rho+a) f\bigg(p +\frac{\sqrt{2\lambda
m}}{\rho+a}\bigg) + (\rho +a) g\bigg(p
-\frac{\sqrt{2\lambda m}}{\rho+a}\bigg) \ .
\label{eq:103plus3}
\eeq
 The physical meaning of this general system is not
clear.

We conclude the discussion on the one-dimensional nonrelativistic
Chaplygin gas by presenting a set of moving solutions, which
are the Galileo boosts of the static solutions (\ref{eq:65plus4}),
(\ref{eq:65plus5}) in Section~III\null. These read
\beq
p = p(x-ut) ~,~~ \rho = \frac{\sqrt 2 \lambda m}{|p-mu|}
\label{eq:103plus4}
\eeq
with $p(x-ut)$ an arbitrary function of $x-ut$ (provided it
never equals $mu$). Clearly this is a constant-profile solution
moving with a velocity $u$.

For the relativistic Born-Infeld system, the Hamiltonian
density (\ref{eq:70}) is
\beq
\Hh = \sqrt{\rho^2c^2+a^2}\sqrt{m^2c^2 + p^2} \ .
\label{eq:104}
\eeq
 Relation (\ref{eq:194}) for the conserved integrals
reads
\beq m^2 (\rho^2c^2+a^2)^2 \Ff_{\rho \rho} = a^2 (m^2c^2 +
p^2 )^2 \Ff_{p p} \ .
\label{eq:105}
\eeq 
This can be solved by separation of variables. We
prefer again, however, to define Riemann coordinates and
give the  solution in terms of arbitrary functions of one
variable, just as in the nonrelativistic case. The relevant
combinations here are
\beq R_\pm = \phi_\rho \pm \phi_p\qquad \phi_\rho =
\arctan\frac{\rho c} {a}\qquad  \phi_p = \arctan\frac{p}{mc}
\label{eq:106}
\eeq and the equations of motion read, in terms of these,
\beq
\dot{R}_\pm = \mp c (\cos R_\mp) R'_\pm
\label{eq:106plusone}
\eeq
 while the general solution to (\ref{eq:105}) is
\beq
\Ff = \frac{mca}{\cos \phi_\rho \cos \phi_p} \; \Big[ f(
\phi_\rho + \phi_p ) + g( \phi_\rho - \phi_p ) \Big] \ .
\label{eq:107}
\eeq 
By choosing, as before, simple monomials for $f$ and
$g$ we get the tower of conserved quantities
\beq I_n^\pm = mca \int \rmd x\, \frac{( \phi_\rho \pm \phi_p
)^n}{\cos \phi_\rho \cos \phi_p} \ .
\label{eq:108}
\eeq
 The Hamiltonian is included as $H=I_0^\pm$. The
momentum, on the other hand, is an infinite series in the
above integrals, requiring
$f$ and $g$ to be exponential functions. We can give an
alternative tower of complex conserved integrals involving
only algebraic functions of $p$ and~$\rho$:
\beq {\tilde I}_n^\pm = \int \rmd x\, \sqrt{\rho^2c^2+a^2}^{\, 1-n}
\sqrt{m^2c^2 + p^2}^{\, 1-n} (\rho c \pm ia)^n (p+imc)^n \ .
\label{eq:109}
\eeq Then the Hamiltonian is $H={\tilde I}_0^\pm$ while the
total momentum, total particle number and integral of the
local momentum are contained in the real and imaginary
parts of $I_1^+$ and $I_1^-$ (as well as a trivial constant). It
can be checked that the integrals (\ref{eq:109}) go over to
the nonrelativistic ones (\ref{eq:198})  in the limit $c \to
\infty$ upon proper rescaling.

In the relativistic model $\rho$ need not be constrained to be positive
(negative $\rho$ could be interpreted as antiparticle density). The
transformation $p \to -p$, $\rho \to -\rho$ is a symmetry and can be
interpreted as charge conjuguation. Further, $p$ and $\rho$ appear in 
an equivalent way. As a result, this theory enjoys a 
duality transformation:
\beq
\rho \to \pm\frac{a}{mc^2} p \qquad  p \to \pm\frac{mc^2}{a}
\rho
\label{eq:110}
\eeq 
Under the above, both the canonical structure and the
Hamiltonian remain invariant. Solutions are mapped in
general to new solutions. Note that the nonrelativistic limit
is mapped to the ultra-relativistic one under duality.
Self-dual solutions $\rho = \pm\frac{a}{mc^2} p$ satisfy
\beq
\dot{\rho} = \mp c \rho'
\label{eq:111}
\eeq and are, therefore, the chiral relativistic solutions that
were presented at the end of Section~III\null.  In the self-dual
case, when $p$ is eliminated from the canonical 1-form and from
the Hamiltonian, one arrives at an action for $\rho$, which
coincides (apart from irrelevant constants) with the
self-dual action, constructed some time ago~\cite{ref:18}:
\beqar
\biggl\{
\half &\displaystyle\int& \rmd x\,\rmd y\, \dot\rho(x)\epsilon (x-y)
p(y)\,\rmd t  - \int dx\,\sqrt{\rho^2c^2+a^2} \sqrt{m^2c^2+p^2}\, \rmd t
\bigg\}\biggr|_{p=\frac{mc^2}a \rho} \nonumber\\
= \frac{2mc^2}a \biggl\{ \quar &\displaystyle\int& \rmd x\,\rmd y\,
\dot\rho(x)\epsilon (x-y)
\rho(y)\,\rmd t  
- \frac c2 \int \rmd x\,\Bigl( \rho^2(x) + \frac{a^2}{c^2}\Bigr)\,
\rmd t\biggr\}\ .\label{eq:new5.29}
\eeqar

A set of constant-profile solutions can be found by Lorentz-boosting
the static solution (\ref{eq:93}). Their most general form is
\beq
p = p(x-ut) ~,\quad \rho = \frac{a}{c} \frac{up \pm mc \sqrt{
c^2 - u^2}}{\left| p\sqrt{c^2 - u^2} \mp mcu \right|}
\label{eq:109plus1}
\eeq
with $p(x-ut)$, again, a general function. Note that the two choices
of sign in the above formula are related by charge conjuguation.
In the extreme relativistic case $u \to c$ this solution goes over to
the chiral relativistic solution (\ref{eq:81}). The set of these
solutions is closed under duality transformations.

We shall conclude by presenting an explicit mapping
between the relativistic and the nonrelativistic theories in
one dimension, which demonstrates their kinematical
equivalence. Note that the relativistic equations of motion
(\ref{eq:106plusone}) become, in terms of $\cos R_\pm$,
\beq
\partial_t \cos R_\pm = \mp c (\cos R_\mp)
\partial_x \cos R_\pm\ .
\label{eq:113}
\eeq These are essentially identical to the equations of
motion for the nonrelativistic Riemann coordinates
(\ref{eq:100}). In fact, putting
\beq {\bar R}_\pm = \pm mc \cos R_\pm
\label{eq:114}
\eeq we see that the ${\bar R}_\pm$ obey the
nonrelativistic equations (\ref{eq:100}). Expressing ${\bar
R}_\pm$ and $R_\pm$ in terms of the corresponding
nonrelativistic and relativistic variables produces a mapping
between the two sets. Call $p_{NR}$, $\rho_{NR}$ and
$p_R$,
$\rho_R$ the  local momentum and density of the
nonrelativistic and the relativistic theory, respectively. Then
the mapping is
\beq
\rho_{NR} = \frac{\sqrt{2\lambda m}}{a m^2 c^2} {\cal H}_R
~,\quad p_{NR} = m c^2 \frac{\rho_R p_R}{{\cal H}_R}
\label{eq:115}
\eeq where ${\cal H}_R = \sqrt{\rho_R^2c^2+a^2}
\sqrt{m^2c^2 + p_R^2}$ is the relativistic Hamiltonian
density. As can be checked by direct algebra, this maps the
relativistic equations of motion to the nonrelativistic  ones.
Since the combinations of
$p_R$ and $\rho_R$ that appear in (\ref{eq:115}) are
duality-invariant, the mapping of solutions is two-to-one.
Note that the constant-profile relativistic solutions 
(\ref{eq:109plus1}) are mapped to the corresponding 
nonrelativistic ones (\ref{eq:103plus4}).

We stress that the transformation (\ref{eq:115}) is not
canonical, since it does not preserve the Poisson brackets.
Accordingly, it does not map the
relativistic Hamiltonian into the nonrelativistic one.  This is a
manifestation of the bi-Hamiltonian structure of these systems, since
there are now two pairs of Hamiltonian and canonical structure (the
standard nonrelativistic one and the one obtained through this  mapping)
that lead to the same nonrelativistic equations of motion.

We note that a similar mapping between the nonrelativistic
system and the relativistic one in light-cone coordinates was
presented previously by Verosky~\cite{ref:19}. The mapping
(\ref{eq:115}), then, can be considered as the analog of
Verosky's transformation for the Lorentz system, although it
cannot be obtained from it in any straightforward way.

The existence of an infinite set of constants of motion for
both the nonrelativistic Chaplygin gas, (\ref{eq:197}) and
(\ref{eq:198}), as well as for the relativistic Born-Infeld
model, (\ref{eq:107})--(\ref{eq:109}), signals the complete
integrability of these theories.  The actual integration of the
equations of motion can be carried out only indirectly.  First
of all, since both these $d=1$ models descend, via alternate
parametrization choices, from the Nambu-Goto string
(1-brane) on the plane (2-space), the explicit integration of
the latter \cite{ref:4} allows presenting solutions of the
former two in terms of two arbitrary functions.
Alternatively, the Chaplygin gas equations can be combined,
after a Legendre transformation, into a linear, second-order
partial differential equation \cite{ref:9}, whose general
solution, in terms of two arbitrary functions, is known
explicitly~\cite{ref:20}.  The Born-Infeld solution can then
be constructed with the help of transformations described in
Sections~IV and V.

\renewcommand{\baselinestretch}{1}


\begin{thebibliography}{99}
\bibitem{ref:15} This result was established with a
hodographic transformation for
$d=2$ by Bordemann and Hoppe (second paper in Ref.~[2]);
see also Hoppe, Ref.~[14].
\end{thebibliography}
\end{document}